\documentclass[%
 reprint,
superscriptaddress,
 amsmath,amssymb,
 aps,
]{revtex4-2}

\usepackage{graphicx}
\usepackage{dcolumn}
\usepackage{bm,}
\usepackage{dcolumn}
\usepackage{array,multirow}
\usepackage{hyperref}
\usepackage{xcolor}
\usepackage{braket}
\usepackage{subfiles}

\begin{document}

\preprint{APS/123-QED}

\title{Electrical tuning of moir\'e excitons in MoSe$_2$ bilayers}

\author{Joakim Hagel}
  \email{joakim.hagel@chalmers.se}
  \affiliation{%
Department of Physics, Chalmers University of Technology, 412 96 Gothenburg, Sweden\\
}%
\author{Samuel Brem}%
\affiliation{%
 Department of Physics, Philipps University of Marburg, 35037 Marburg, Germany\\
}%
  \author{Ermin Malic}%
  \affiliation{%
 Department of Physics, Philipps University of Marburg, 35037 Marburg, Germany\\
}%
\affiliation{%
Department of Physics, Chalmers University of Technology, 412 96 Gothenburg, Sweden\\
}%

\date{\today}

\begin{abstract}
Recent advances in the field of vertically stacked 2D materials have revealed a rich exciton landscape. In particular, it has been demonstrated that out-of-plane electrical fields can be used to tune the spectral position of spatially separated interlayer excitons. Other studies have shown that there is a strong hybridization of exciton states, resulting from the mixing of electronic states in both layers. However, the connection between the twist-angle dependent hybridization and field-induced energy shifts has remained in the dark. Here, we investigate on a microscopic footing the interplay of electrical and twist-angle tuning of moir\'e excitons in MoSe$_2$ homobilayers. We reveal distinct energy regions in PL spectra that are clearly dominated by either intralayer or interlayer excitons, or even dark excitons. Consequently, we predict twist-angle-dependent critical electrical fields at which the material is being transformed from a direct into an indirect semiconductor. Our work provides new microscopic insights into experimentally accessible knobs to significantly tune the moir\'e exciton physics in atomically thin nanomaterials. 
\end{abstract}
\maketitle

\begin{figure}[t!]
\hspace*{-0.5cm}  
\includegraphics[width=0.85\columnwidth]{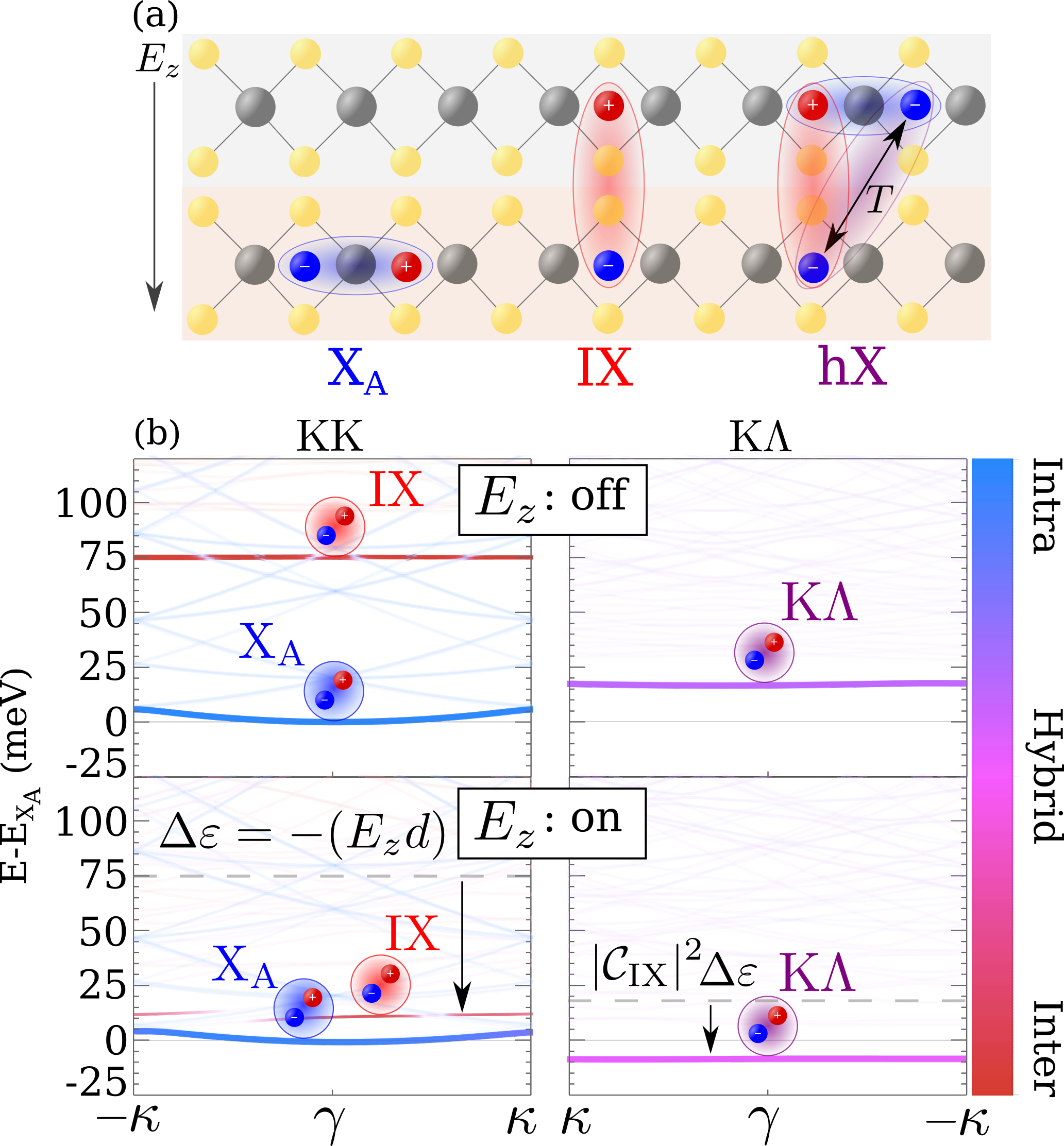}
\caption{\label{fig:1}\textbf{(a)} Real space schematic for different exciton species in TMD bilayers including the intralayer exciton X$_{\text{A}}$, the interlayer exciton IX, and the hybrid exciton hX stemming from interlayer tunneling $T$. \textbf{(b)} Exciton band structure for R-type stacked MoSe$_2$ homobilyer for the bright KK (left) and dark K$\Lambda$ excitons (right). Multiple moir\'e subbands are shown in faded colors. Only the interlayer and the hybrid exciton exhibiting a dipole moment become red-shifted in presence of an electrical field (lower panel). Dashed lines indicate the original position of the corresponding excitons without an electrical field. The calculation is performed for a twist angle of $\theta=2^{\circ}$ and an electrical field of $E_z=0.1$ V/nm.}
\end{figure}

The realization of atomically thin nanomaterials, such as graphene and monolayers of transition metal dichalcogenides (TMDs), has opened up a new platform to study many-particle quantum phenomena \cite{ares2021recent,tong2017topological,cao2018unconventional,sung2020broken}. Optical properties are of particular interest for TMDs, where strongly bound electron-hole pairs can form. They give rise to exciton features, which fundamentally change the optical response of the material \cite{mueller2018exciton,splendiani2010emerging,he2014tightly,PhysRevLett.113.076802,wang2018colloquium}.
By vertically stacking two TMD monolayers, long-lived interlayer excitons become possible, thus allowing for both the intralayer A exciton resonance (X$_\text{A}$) and the spatially separated interlayer exciton resonance (IX) in optical spectra \cite{rivera2015observation,geim2013van,liao2019van,kunstmann2018momentum,miller2017long,jin2018ultrafast,gillen2018interlayer,merkl2019ultrafast,kunstmann2018momentum,rivera2018interlayer}, cf. \autoref{fig:1}(a). Furthermore, the exciton can become strongly hybridized (hX in \autoref{fig:1}(a)) via interlayer carrier tunneling, and thus carry both an inter- and an intralayer component \cite{alexeev2019resonantly,gerber2019interlayer,D0NR02160A,PhysRevResearch.3.043217}. One example of strongly hybridized states states are momentum-dark K$\Lambda$ excitons, where the hole is located at the K valley and the electron resides in the $\Lambda$ valley in the Brillouine zone. Due to the strong wavefunction overlap at the $\Lambda$ valley, the electron is delocalized over both layers \cite{D0NR02160A,PhysRevResearch.3.043217}.

Due to the charge separation of the interlayer exciton, it exhibits an out-of-plane dipole moment, allowing for external tuning via an electrical field for both the pure interlayer exciton as well as the hybrid exciton \cite{jiang2021interlayer,liu2020electrically}. Additionally, one can twist the TMD monolayers with respect to each other, which introduces a spatially periodic potential. At small twist angles, this moir\'e potential was shown to give rise to a flat band structure that traps excitons in real space \cite{seyler2019signatures,tran2019evidence,Tong_2020,yu2017moire,zhang2018moire,merkl2020twist,kiemle2020control,brem2020tunable, D0NR02160A}.
Previous experimental studies have shown that one can efficiently tune the optical properties of TMDs by applying an external out-of-plane electrical field \cite{huang2022spatially,altaiary2022electrically,leisgang2020giant,peimyoo2021electrical,SponfeldnerMoS2,wang2018electrical,PhysRevB.84.205325,wu2013electrical}, however the interplay with the twist-angle-dependent tuning of exciton hybridization has not been explored microscopically. In this work, we investigate this interplay with a fully microscopic and material-specific theory based on the density-matrix formalism combined with first-principle calculations \cite{brem2020tunable,D0NR02160A,PhysRevResearch.3.043217}. 
We predict twist-angle dependent critical electrical fields at which the MoSe$_2$ bilayer becomes an indirect semiconductor, where the dark K$\Lambda$ exciton is clearly the energetically lowest state. The low-temperature PL spectrum is then dominated by the phonon sidebands of this dark state. Depending on the electrical field and the twist-angle we find spectral regions, where either the intralayer X$_{\text{A}}$, the dark K$\Lambda$ or the interlayer exciton IX are clearly dominating the PL spectrum. 

\textbf{Theoretical approach:} To obtain a microscopic access to the optical response of moir\'e excitons, we first set up an excitonic Hamilton operator \cite{katsch2018theory,D0NR02160A,brem2020tunable}, using monolayer eigenstates as basis,
\begin{equation}\label{eq:ExcitonHam}
\begin{split}
    H_0&=\sum_{L\bm{Q}\bm{\xi}}E^{\bm{\xi}}_{L\bm{Q}}X^{\bm{\xi}\dagger}_{L,\bm{Q}}X^{\bm{\xi}}_{L,\bm{Q}}+\sum_{L\bm{Q}\bm{\xi}\bm{g}}V^{\bm{\xi}}_{L\bm{g}}X^{\bm{\xi}\dagger}_{L,\bm{Q}+\bm{g}}X^{\bm{\xi}}_{L,\bm{Q}}\\
    &+\sum_{LL^{\prime}\bm{Q}\bm{\xi}\bm{g}}T^{\bm{\xi}}_{LL^{\prime}\bm{g}}X^{\bm{\xi}\dagger}_{L,\bm{Q}+\bm{g}}X^{\bm{\xi}}_{L^{\prime},\bm{Q}}+h.c,
    \end{split}
\end{equation}
where $L=(l_e,l_h)$ is the compound index for electron/hole layer $l_{e(h)}$, $\bm{Q}$ the center-of-mass momentum, and $\bm{\xi}=(\xi_e,\xi_h)$ the compound valley index allowing for both bright and momentum-dark excitons \cite{PhysRevMaterials.2.014002,berghauser2018mapping}. Furthermore, $\bm{g}$ is the moir\'e lattice vector of the mini-Brillion zone (mBZ)\cite{D0NR02160A,brem2020tunable}, while $X^{(\dagger)}$ are annihilation (creation) operators for layer localized excitons. The first term in \autoref{eq:ExcitonHam} is the kinetic energy of excitons and includes their center-of-mass exciton dispersion as obtained by solving the bilayer Wannier equation \cite{ovesen2019interlayer}. The following two terms are twist-angle dependent, where $V^{\bm{\xi}}_{L\bm{g}}$ is the periodic polarization-induced interlayer alignment shift (here referred to as moir\'e-induced shift) taking into account the periodic modulation of interlayer excitons \cite{brem2020tunable,tong2020interferences}. Furthermore, $T^{\bm{\xi}}_{LL^{\prime}\bm{g}}$ is the interlayer tunneling, taking into account the hybridization of electronic states in both layers \cite{D0NR02160A,PhysRevResearch.3.043217}. The moir\'e-induced shift and the hybridization both give rise to a periodic moir\'e potential that is present in TMD bilayers \cite{MoireExcitonsLinderalv}. A complete derivation of the Hamiltonian and the appearing matrix elements is provided in the supplementary material. 

To take into account the periodicity of the moir\'e superlattice, we apply a zonefolding approach and consequently restrict the summation over $\bm{Q}$ to the first mBZ.\cite{brem2020tunable,D0NR02160A}. The moir\'e exciton band structure can be calculated by transforming the zone-folded Hamiltonian into a moir\'e exciton basis with the new exciton creation operators $Y^{\dagger}_{\bm{\xi}\eta\bm{Q}}=\sum_{\bm{g}L}\mathcal{C}^{\bm{\xi}\eta*}_{L\bm{g}}(\bm{Q})X^{\bm{\xi}\dagger}_{L,\bm{Q}+\bm{g}}$. The mixing coefficients $\mathcal{C}^{\bm{\xi}\eta}_{L\bm{g}}(\bm{Q})$ contain the relative contribution of intra- and interlayer excitons as well the the contribution of each subband that emerges from twisting the structure. Here, $\eta$ is the new moir\'e exciton band index. This transformation leads to the moir\'e exciton eigenvalue equation
\begin{equation}\label{eq:eigenvalue}
    \begin{split}
        E^{\bm{\xi}}_{L\bm{Q+g}}\mathcal{C}^{\bm{\xi}\eta}_{L\bm{g}}(\bm{Q})+\sum_{\bm{g}^{\prime}}V^{\bm{\xi}}_L(\bm{g},\bm{g}^{\prime})\mathcal{C}^{\bm{\xi}\eta}_{L\bm{g}^{\prime}}(\bm{Q})\\
        +\sum_{L^{\prime}\bm{g}^{\prime}}T^{\bm{\xi}}_{LL^{\prime}}(\bm{g},\bm{g}^{\prime})\mathcal{C}^{\bm{\xi}\eta}_{L^{\prime}\bm{g}^{\prime}}(\bm{Q})=\mathcal{E}^{\bm{\xi}}_{\eta\bm{Q}}\mathcal{C}^{\bm{\xi}\eta}_{L\bm{g}}(\bm{Q}).
    \end{split}   
\end{equation}
Solving this eigenvalue problem numerically gives a microscopic access to the hybrid moir\'e exciton energies $\mathcal{E}^{\bm{\xi}}_{\eta\bm{Q}}$ and the corresponding wave functions. 

Figure \ref{fig:1}(b) shows the corresponding exciton band structure for the R-type stacked hBN-encapsulated MoSe$_2$ bilayer for a twist angle of $\theta=2^{\circ}$. The top panel shows the case without an electrical field and illustrates the bright KK excitons on the left and the dark K$\Lambda$ excitons on the right. We find that at $\theta=2^{\circ}$, the A exciton is the energetically lowest lying state directly followed by the strongly hybridized dark K$\Lambda$ exciton. Further up in the band structure, the interlayer IX exciton can be found. We predict that both the IX and the K$\Lambda$ exciton exhibit very flat bands. This can be ascribed to the moir\'e-induced shift $V^{\bm{\xi}}_L(\bm{g})$ that traps interlayer excitons in real space \cite{brem2020tunable}. The same holds for K$\Lambda$ excitons on top of a strong carrier tunneling, which is also periodic over the superlattice and contributes to the flatness of the band \cite{D0NR02160A,PhysRevResearch.3.043217}. The efficient tunneling of electrons at the $\Lambda$ valley ($t^{\text{max}}_c=195$ meV \cite{PhysRevResearch.3.043217}) is responsible for the pronounced hybrid nature of this exciton, which exhibits nearly a 50/50 contribution from both its intra- and interlayer exciton components (cf. the color gradient in Fig. \ref{fig:1}(b). Furthermore, the solution of the moir\'e eigenvalue equation \ref{eq:eigenvalue} also reveals the formation of multiple subbands emerging from the introduction of the twist angle and the resulting superlattice. The subbands stemming from the intralayer exciton ($X_\text{A}$) can then hybridize with the energetically higher interlayer IX exciton, cf. the avoided crossings in the interlayer exciton in Fig. \ref{fig:1}(b).

\textbf{Moir\'e excitons in electrical fields:}
In presence of an out-of-plane electrical field, excitons become shifted in energy proportional to their dipole moment. The exciton-field interaction Hamiltonian reads
\begin{equation}
    H_{X-l}=-\sum_{\bm{\xi}\bm{Q}L}d_LE_zX^{\bm{\xi}\dagger}_{\bm{Q},L}X^{\bm{\xi}}_{\bm{Q},L}\,\,\,.
\end{equation}
with $d_L$ as the dipole moment and $E_z$ as the out-of-plane electrical field. Since this Hamiltonian is simply a renormalization of the potential energy of excitons, it can easily be incorporated into the first term of \autoref{eq:ExcitonHam}. Solving again the eigenvalue equation (\autoref{eq:eigenvalue}), we can now calculate the moir\'e exciton band structure in presence of an electrical field, cf. the lower panel in \autoref{fig:1}(b) for the MoSe$_2$ bilayer at $\theta=2^{\circ}$ and $E_z=0.1$ V/nm. 

We predict that the interlayer exciton IX red-shifts in energy by $\Delta\varepsilon=-(E_zd)$ and becomes very close to the intralayer exciton X$_{\text{A}}$. Interestingly, we show clear avoided crossings with the higher-lying moir\'e subbands of X$_{\text{A}}$. As the states are nearly degenerate even the weak tunneling of carriers at the K point has a significant effect on the exciton band structure. There is also a field-induced red-shift of the dark K$\Lambda$ exciton as it has an interlayer component (given by $|\mathcal{C}_{\text{IX}}|^2\Delta\varepsilon$) due to the strong hybridization. As a consequence, the K$\Lambda$ exciton becomes the energetically lowest state at the considered twist angle of $\theta=2^{\circ}$, cf. the lower panel of Fig. \ref{fig:1}(b).

\begin{figure}[t!]
\hspace*{-0.5cm}  
\includegraphics[width=\columnwidth]{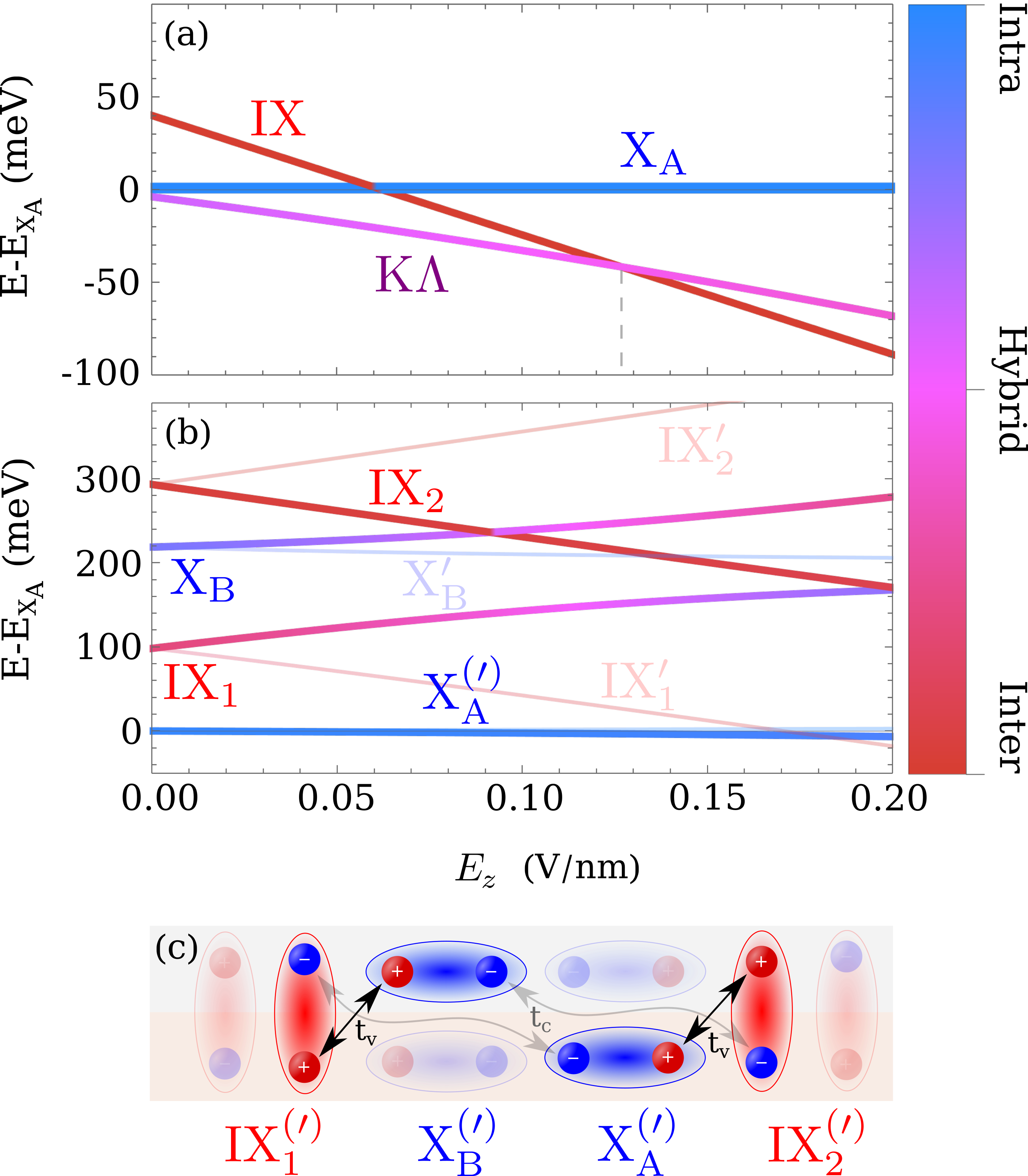}
\caption{\label{fig:2}\textbf{(a)} Electrical field-induced tuning of exciton energy for the untwisted $R^M_h$-stacked MoSe$_2$ bilayer. The color gradient denotes the degree of hybridization between the intra- and interlayer excitons. The strongly hybridized K$\Lambda$ exciton is the energetically lowest state in a certain $E_z$-range, while at even stronger fields the interlayer IX exciton becomes the lowest state. \textbf{(b)} Exciton band structure for $H^h_h$-stacked MoSe$_2$ bilayer. There is a strong hybridization between the field-shifted interlayer exciton IX$_1$ and the B exciton X$_\text{B}$ (change of color at the avoided crossing). Shaded lines indicate the degenerate exciton species X$^{\prime}$ that have the reversed dipole moment. \textbf{(c)} Schematic for the tunneling processes that couple different excitons in (b). IX$_1$ and X$_\text{B}$ couple with a strong hole tunneling $t_v$, while the hybridization is zero between IX$_2$ and X$_\text{B}$ due to symmetry constraints. 
Shaded excitons correspond to X$^{\prime}$ states that do not couple with X excitons via tunneling.
}
\end{figure}

Now, we investigate how the exciton bands can be tuned with the electrical field in the untwisted $R^M_h$-stacked MoSe$_2$ bilayer, cf. \autoref{fig:2}(a). At very low electrical fields, the dark K$\Lambda$ exciton is found to be the lowest state, i.e. the untwisted MoSe$_2$ bilayer is an indirect semiconductor, however note that the bright X$_{\text{A}}$ exciton is very close in energy. By introducing a twist angle (as in \autoref{fig:1}), the K$\Lambda$ exciton lies above the bright exciton due to the reduced hybridization with the twist angle (caused by the momentum mismatch in the rotated Brilloine zones) \cite{D0NR02160A,merkl2020twist}. When increasing the electrical field, we observe a critical value of $E_z\approx0.14$ V/nm at which the interlayer exciton IX becomes the lowest lying state, cf. \autoref{fig:2}(a). This is due to its larger effective dipole moment compared to the K$\Lambda$ exciton (that is only partially interlayer-like) resulting in a steeper field-induced red-shift of the IX exciton. As the intralayer exciton X$_{\text{A}}$ does not have a dipole moment and the hybridization is very weak at the K point, it remains unchanged in presence of an electrical field.

We have performed the same calculation for untwisted $H^h_h$-stacked MoSe$_2$ bilayers, cf. \autoref{fig:2}(b). Due to the 180$^{\circ}$-rotation of one of the layers in this stacking, the K$^{\prime}$ valley is now on top of the K valley, which inverts the spin-orbit coupling in one of the layers. This fundamentally changes the tunneling channels in comparison to R-type stacking. Here, we have the A exciton X$_{\text{A}}$ in one layer coupling to the B exciton X$_{\text{B}}$ in the other layer via the two interlayer states IX$_1$ and IX$_{2}$. There is also a degenerate exciton species X$^{\prime}_{\text{A}}$ (X$^{\prime}_{\text{b}}$) with the reversed layer configuration that couple via their respective interlayer excitons IX$^{\prime}_1$ (IX$^{\prime}_{2}$). These interlayer excitons have the reversed dipole moment in comparison to IX$_1$ (IX$_{2}$), cf. \autoref{fig:2}(c). Note that X and X$^{\prime}$ states do not couple to each other via interlayer tunneling as these excitons are separated by large momentum in the Brillouin zone. These considerations plus the slight reduction of the interlayer distance for $H_h^h$ stacking (which increases the strength of the hole tunneling \cite{PhysRevResearch.3.043217}) results in a significant hybridization between the energetically lowest interlayer exciton IX$_1$ and the B exciton X$_{\text{B}}$. The higher interlayer exciton IX$_2$ instead couples to the B exciton via electron tunneling (cf. \autoref{fig:2}(c)), which is zero at this stacking due to the C3-symmetry at the K point (note that this symmetry only holds at $\theta=0^{\circ}$)\cite{PhysRevResearch.3.043217}. By increasing the electrical field we find a slightly non-linear increase of X$_B$ and IX$_1$ excitons. This can be ascribed to the fact that the tunneling strength between these excitons is in the same range as the shift induced by the electrical field. This allows for strong tuning of the mixing coefficients $\mathcal{C}^{\bm{\xi}\eta}_{L\bm{g}}(\bm{Q})$. The color gradient encodes the degree of hybridization and shows nicely the strong hybridization between X$_B$ and IX$_1$ in the range around $E_z\approx0.1$ V/nm (pink color). Furthermore, due to the reversed dipole moment of the degenerate exciton species X$^{\prime}$, the interlayer excitons IX$_1^{\prime}$ (IX$_2^{\prime}$) are shifted in the opposite direction, as can be seen from the shaded lines in \autoref{fig:2}(b). 

\textbf{PL spectra of R-stacked MoSe$_2$ bilayers:}
Having obtained microscopic insights into the tunability of the exciton landscape with the electrical field in MoSe$_2$ bilayers, we now focus on the change of their optical response. We first calculate the PL spectrum of R-stacked bilayers and we explicitly consider also indirect phonon-driven exciton recombination processes \cite{D0NR02160A,PLBrem}. Here, momentum-dark states can become visible via scattering with a phonon to a virtual state within the light cone and resulting in the formation of phonon sidebands in PL spectra. These are expected to play an important role for R-stacked MoSe$_2$ bilayers, as here the momentum-dark K$\Lambda$ excitons are the lowest states in a large range of electrical fields (cf. \autoref{fig:2}(a)). 

\begin{figure}[t!]
\hspace*{-0.0cm}  
\includegraphics[width=\columnwidth]{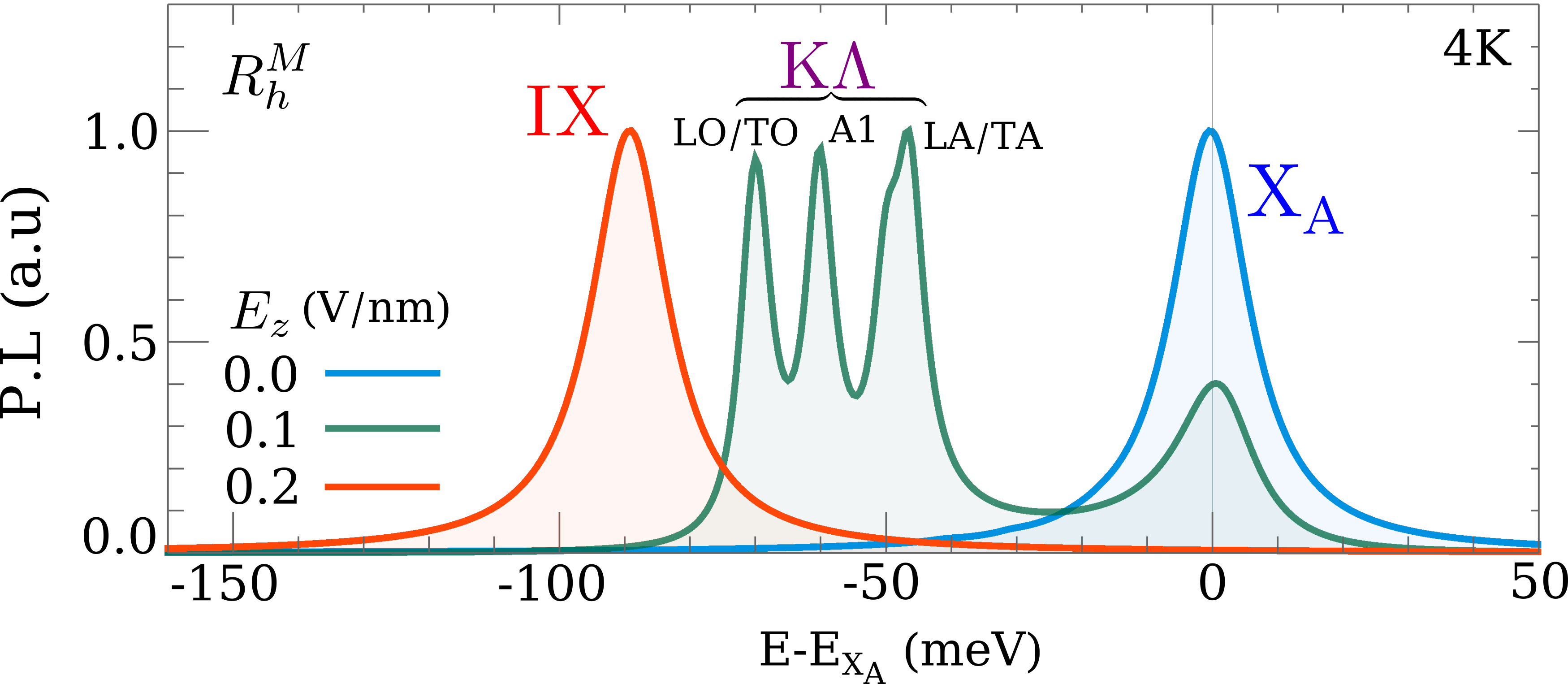}
\caption{\label{fig:3}Photoluminescence spectrum for $R^M_h$-stacked MoSe$_2$ bilayers for three different electrical fields at 4K. Without the field, the A exciton clearly dominates the PL and no other features are visible. When applying an out-of-plane electrical field ($E_z=0.1$ (V/nm)), the dark K$\Lambda$ exciton is significantly red-shifted and its occupation becomes so large, that the phonon sidebands from this state dominate the PL. For even larger fields ($E_z=0.2$ (V/nm)), the strong red-shift of the interlayer exciton makes it to the lowest state and its emission dominates the PL. }
\end{figure}

Figure \ref{fig:3} shows the PL spectrum for three different values of the electrical field $E_z$ at $T=4$ K. Without a field, the A exciton clearly dominates the PL. Even though the K$\Lambda$ state is predicted to be slightly the lowest state, it is only so by a couple of meV (Fig. \ref{fig:2}(a)). We do not observe here any phonon sidebands, as the phonon-assisted recombination is a higher-order process with a much smaller probability compared to the direct recombination. The predicted close proximity of the K$\Lambda$ exciton to the bright A exciton according to the solution of the moir\'e eigenvalue equation agrees well with previous DFT studies \cite{deilmann2019finite} as well as with experimental data showing the dominance of the A exciton resonance \cite{Liu_2015}.

By applying an electrical field we decrease the energy of the K$\Lambda$ exciton pushing it far below the bright exciton. As a direct consequence, this dark state carries by far the largest occupation, which makes the phonon-assisted recombination dominate over the direct recombination process. At an electrical field of $E_z=0.1$ V/nm, we find prominent phonon sidebands from K$\Lambda$ excitons, cf. Fig. \ref{fig:3}. The three distinct peaks stem from different phonon modes including the longitudinal/transverse optical modes (LO/TO), the A1 out-of-plane mode and the longitudinal/transverse acoustical modes (LA/TA). Increasing the electrical field further shifts the interlayer exciton IX (with the largest dipole moment and thus the steepest shift with the field) to the energetically lowest state (\autoref{fig:2}(a)). Consequently, the radiative recombination of the IX dominates the PL spectrum for large fields, cf. Fig. \ref{fig:3}. 

\begin{figure}[t!]
\hspace*{-0.0cm}  
\includegraphics[width=\columnwidth]{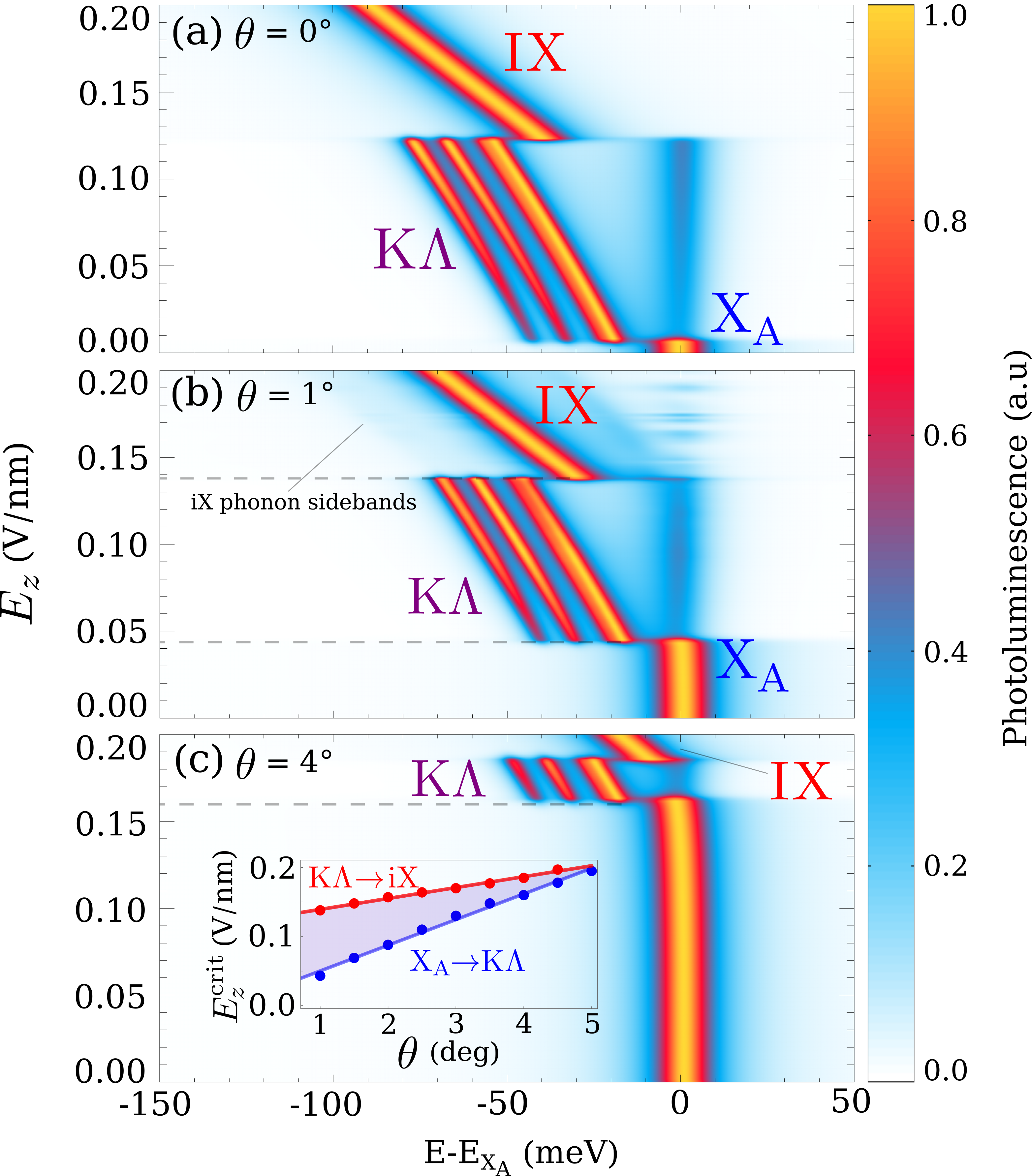}
\caption{\label{fig:4}Photoluminescence spectra as a function of electrical field strength at the twist angles of (a) $\theta=0^{\circ}$, (b) $\theta=1^{\circ}$ and (c) $\theta=4^{\circ}$ at 4K. We observe three distinct energy regions, which are dominated by the A exciton X$_{\text{A}}$, the phonon sidebands of the dark K$\Lambda$ exciton and the interlayer exciton IX, respectively. The inset in (c) shows the critical electrical field strength $E^{\text{crit}}_z$ as a function of the twist angle for the transition between different spectral regions (X$_{\text{A}}$ to K$\Lambda$ and X$_{\text{A}}$ to IX-dominated region). The shaded area indicates the range, where the dark K$\Lambda$ excitons dominate the PL.}
\end{figure}

In a similar fashion as in \autoref{fig:2}, we now investigate how the optical response of untwisted MoSe$_2$ bilayers evolves when continuously changing the electrical field, cf. \autoref{fig:4}(a). At zero electrical field, the A exciton dominates the PL, however we observe a drastic change in the spectrum with just a slight increase of the field. Here, the K$\Lambda$ exciton is sufficiently red-shifted to gain enough occupation to give rise to the clearly visible phonon sidebands. Due to the close proximity of the K$\Lambda$ and the A exciton and the low temperature of $T=4$ K, this change occurs rather abruptly. The phonon sidebands shift linearly with the electrical field due to the interlayer component of the strongly hybridized K$\Lambda$ exciton. Around $E_z\approx0.125$ V/nm, we predict another drastic change in the PL spectrum. Here, the interlayer exciton IX has become the energetically lowest state as it has a larger dipole moment than the K$\Lambda$ exciton and thus exhibits a steeper shift with the electrical field (\autoref{fig:2}(a)). As the IX state is a bright exciton and recombines directly, we find again a single peak. It is blue-shifted with respect to the phonon sidebands of the K$\Lambda$ exciton, since no phonon energy is involved in this recombination process.

Next we investigate how the field-induced tuning of the PL spectrum changes when introducing a finite twist angle and thus a moir\'e superlattice. For an angle of $\theta=1.0^{\circ}$, we find the qualitatively same drastic changes in the PL spectrum at specific critical field values, cf. Fig. \ref{fig:4}(b). Compared to the untwisted bilayer, there is an increased field range at which the A exciton dominates. The reason for this is an increased K$\Lambda$ exciton energy in a twisted structure without electrical field. As result of the twist angle, the two monolayer band structures are rotated with respect to each other leading to an energetic detuning of hybridizing states. Moreover, the spatially varying stacking sequences result in variations of the interlayer distance, leading to a weaker tunneling than in the optimal $R^M_h$ stacking discussed in Fig. \ref{fig:4}(b). Further side effects leading to weaker hybridization with increasing twist angle have been discussed and experimentally observed in Ref. \cite{merkl2020twist}. 

As a result of the above, it requires a larger electrical field to make this state energetically lowest and dominate the PL spectrum via its phonon sidebands. Furthermore, the rotation of the BZs leads to a energetic minimum of the interlayer exciton IX outside the light cone due to the momentum mismatch between the K points in the different layers. Consequently, there is a higher occupation in the finite-momentum regions of the interlayer exciton valley. This gives rise to some very weak phonon side bands of the IX exciton. The oscillations appearing at the energy of the X$_{\text{A}}$ exciton stem from the hybridization of multiple IX subbands with the A exciton, cf. also the avoided crossings in \autoref{fig:1}(b).

Further increasing the twist angle up to $\theta=4^{\circ}$, the field region with the dominant K$\Lambda$ phonon sidebands has become very small, cf. \autoref{fig:4}(c). Instead, the A exciton nearly dominates the entire energy range. K$\Lambda$ and IX excitons become important only at relatively large fields of $E_z>0.16$V/nm. This can again be ascribed to the decrease in the hybridization of K$\Lambda$ excitons and the increased momentum-indirect character of the IX \cite{brem2020tunable, choi2021twist}, as discussed above. The inset in \autoref{fig:4}(c) describes the twist-angle dependent critical electrical field necessary to give rise to the drastic changes in the PL spectrum from spectral regions with a dominant X$_{\text{A}}$ to K$\Lambda$ (blue) as well as from K$\Lambda$ to IX (red). We find that this critical field changes approximately linearly with the twist angle and that the transition X$_{\text{A}}$ to K$\Lambda$ has a steeper slope. This can be explained by the blue-shift of both excitons when increasing the size of the mBZ. Moreover, the K$\Lambda$ exciton has a smaller effective dipole moment, making its slope less steep compared to the IX exciton. As a result, the K$\Lambda$ dominated electric field region shrinks with increasing twist angle, cf. the color-shaded region in the inset.

\begin{figure}[t!]
\hspace*{-0.5cm}  
\includegraphics[width=\columnwidth]{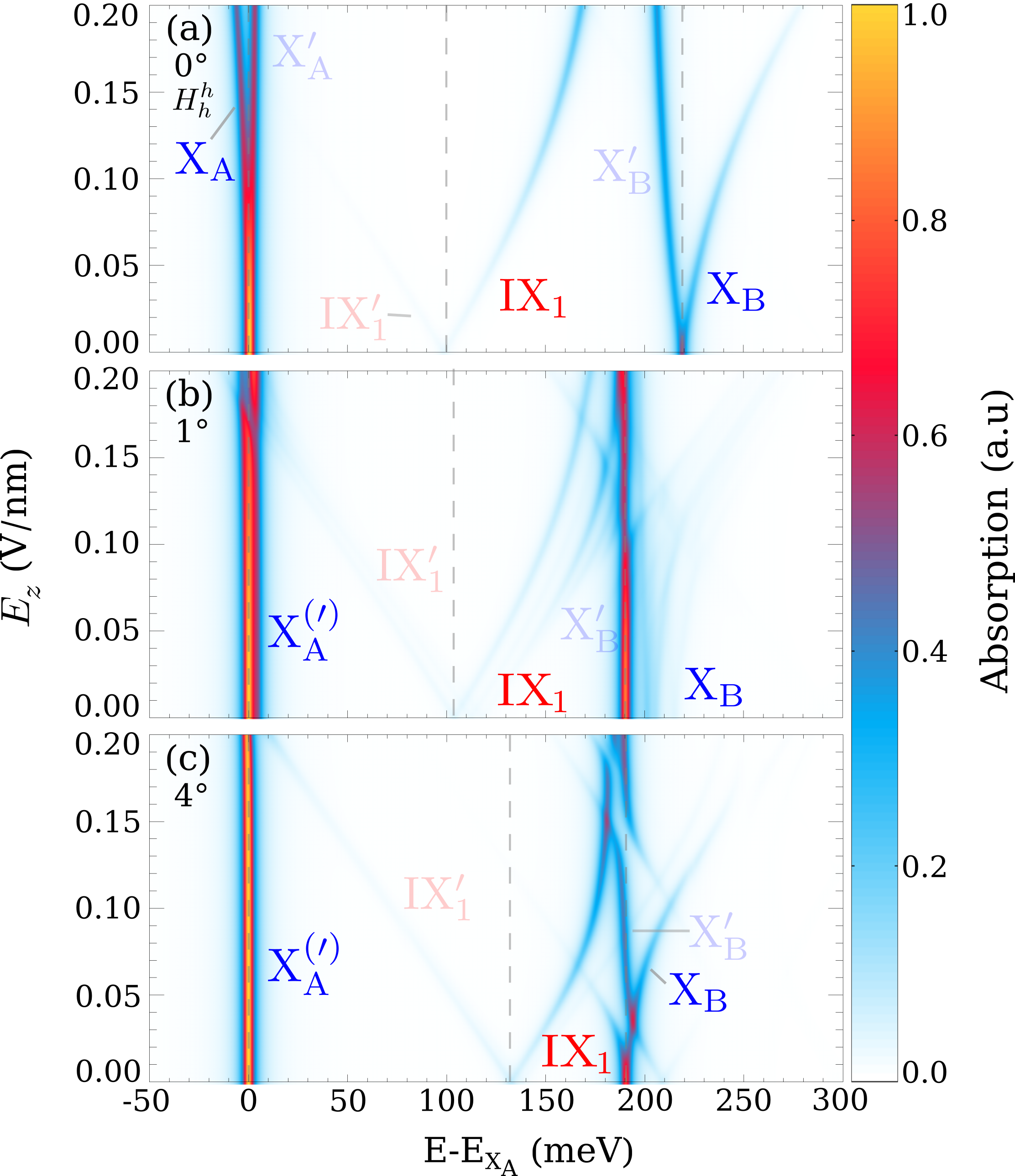}
\caption{\label{fig:5}Absorption spectra for H-type stacked MoSe$_2$ as a function of electrical field $E_z$ at the twist angles of (a) $\theta=0^{\circ}$, (b) $\theta=1^{\circ}$ and (c) $\theta=4^{\circ}$. We observe that the interlayer exciton IX$_1$ carries a considerable oscillator strength and has a pronounced avoided crossing with the B exciton. Dashed lines indicates the position of the peak without electrical field. Note that in contrast X$_2$ is not visible as here the hybridization with the X$_\text{B}$ state is negligible, cf. Fig. \ref{fig:2}(b).}
\end{figure}

\textbf{Absorption spectra of H-stacked bilayers:} Finally, we investigate the H-stacked MoSe$_2$ bilayer. The corresponding exciton band structure has already been discussed in \autoref{fig:2}(b), where it has been shown that the B exciton X$_{\text{B}}$ strongly hybridizes with the interlayer exciton IX$_1$. Since both of these excitons are bright, but not the energetically lowest states, their PL signal is very small. Therefore, we now investigate the field-dependent optical response by studying the optical absorption as illustrated in \autoref{fig:5} for different twist angles. In the untwisted case (\autoref{fig:5}(a)), we observe a very large avoided crossing between IX$_1$ and X$_{\text{B}}$, i.e. both lines become strongly non-linear when they come close to each other. This is due to the large hole tunneling around the K point ($t_v=56$ meV \cite{PhysRevResearch.3.043217}), cf. \autoref{fig:2}(c). As a result, the interlayer exciton IX$_1$ will be strongly hybridized and thus gains a considerable oscillator strength. This explains that the otherwise invisible interlayer exciton can be clearly observed in the absorption spectrum - in good agreement with previous experimental studies \cite{leisgang2020giant,SponfeldnerMoS2,peimyoo2021electrical} and also in accordance with a previous DFT work \cite{deilmann2018interlayer}. Note that the IX$^{\prime}_1$ exciton, i.e the degenerate counterpart of IX$_1$ with reversed dipole moment (dashed line in Fig. \ref{fig:2}(c)) is not visible as it decreases with the electrical field and is thus shifted away from its corresponding B exciton X$^{\prime}_\text{B}$. Consequently, X$^{\prime}_\text{B}$ exciton remains almost unchanged under the electrical field. Furthermore, at very large electrical fields, we observe a small splitting of the two A excitons X$_\text{A}$ and X$^{\prime}_\text{A}$. This is as a consequence of its small interlayer component gained via hybridization with the interlayer exciton IX$_{2}$, which become closer with an increasing electrical field, cf. Fig. \ref{fig:2}(b)). 

Now, we go beyond previous studies and investigate how the optical response changes in H-stacked MoSe$_2$ bilayers, when a twist angle is introduced. We start with a small angle of $\theta=1^{\circ}$ (\autoref{fig:5}(b)). We find a qualitatively similar behavior as in the untwisted case, but with additional moir\'e subbands appearing in the spectrum. The higher subbands of the interlayer exciton are similarly affected by the electrical field and hybridize with the subbands of the B exciton X$_\text{B}$ resulting in multiple avoided crossings. Moreover, also the intralayer B exciton is experiencing a superlattice potential resulting in a peak splitting due to the formation of moir\'e trapped states \cite{brem2020tunable}.
By further increasing the twist angle up to $\theta=4^{\circ}$ (\autoref{fig:5}(c)), most moir\'e subbands become dark due to a vanishing moir\'e induced mixing with lowest (bright) subband \cite{brem2020tunable,fitzgerald2022twist}. Furthermore, we observe that the avoided crossing occurs at smaller values of the electrical field compared to the case of $\theta=1^{\circ}$. This is due to the interlayer exciton IX$_{1}$ being much closer to the B exciton X$_{\text{B}}$ at higher twist angles. The other interlayer exciton IX$^{\prime}_1$ is now more pronounced as it is closer to the B exciton and thus exhibits a stronger mixing/larger oscillator strength. Overall, the efficient hole tunneling between the interlayer exciton and the B exciton leads to a significant hybridization in H-stacked MoSe$_2$ bilayers, brightening up the otherwise invisible interlayer exciton. A twist angle introduces additional avoided crossings due to the hybridization with the appearing moir\'e subbands.

In conclusion, our work provides microscopic insights into the tunablity of moir\'e excitons with the electrical field and the twist angle in atomically thin semiconductors. In particular, we demonstrate the significant impact of the interlayer hybridization when applying an out-of-plane electrical field, predicting distinct spectral regions in PL spectra of R-stacked MoSe$_2$ bilayers, where the intralayer or the interlayer exciton or even the momentum dark K$\Lambda$ exciton are clearly dominant. Consequently, we predict twist-angle-dependent critical electrical fields for the transition of the material from a direct into an indirect semiconductor. Overall, we shed light on microscopic many-particle processes behind the electrical tunability of moir\'e excitons in TMD bilayers, which can guide and trigger future experimental studies.\\

This project has received funding from Deutsche Forschungsgemeinschaft via CRC 1083 (project B09) and the European Unions Horizon 2020 research and innovation programme under grant agreement no. 881603 (Graphene Flagship).

\end{document}


\preprint{APS/123-QED}

\title{Supplementary material: Electrical tuning of moir\'e excitons in MoSe$_2$ bilayers}

\author{Joakim Hagel}
  \email{joakim.hagel@chalmers.se}
  \affiliation{%
Department of Physics, Chalmers University of Technology, 412 96 Gothenburg, Sweden\\
}%
\author{Samuel Brem}%
\affiliation{%
 Department of Physics, Philipps University of Marburg, 35037 Marburg, Germany\\
}%
  \author{Ermin Malic}%
  \affiliation{%
 Department of Physics, Philipps University of Marburg, 35037 Marburg, Germany\\
}%
\affiliation{%
Department of Physics, Chalmers University of Technology, 412 96 Gothenburg, Sweden\\
}%

\date{\today}

\maketitle
\section{Theory}

\subsection{Derivation of the moir\'e exciton Hamiltonian}
A microscopic model for the moir\'e exciton landscape is formulated by first considering a Hamiltonian in second quantization. For this purpose we start in a decoupled monolayer basis \cite{brem2020tunable,D0NR02160A}, where the strong Coulomb interaction between electrons and holes is taken into account by solving the generalized Wannier equation \cite{ovesen2019interlayer}. Modifications to the exciton energies arise from the stacking-dependent polarization-induced alignment shift (which we here refer to as moir\'e induced alignment shift) $V^{\lambda}(\bm{r})$ and the interlayer tunneling $T^{\lambda}(\bm{r})$. The moir\'e induced alignment shift $V^{\lambda}(\bm{r})$ stems from the polarization-induced electrostatic potential between the atoms, consequently shifting the alignment of the two monolayer bandstructures \cite{tong2020interferences,Tong_2020,yu2017moire,brem2020tunable}. This shift varies with the atomic configuration and is thus periodic when introducing a twist angle. The interlayer tunneling $T^{\lambda}(\bm{r})$ instead stems from the overlapping wavefunctions of both layers, which allows for interlayer hybridization between the decoupled monolayers \cite{alexeev2019resonantly,gerber2019interlayer,ruiz2019interlayer,D0NR02160A,PhysRevResearch.3.043217}. When introducing a twist angle, the tunneling between the layers becomes also periodic due to the change in interlayer distance throughout the superlattice and varying overlap geometries of the involved orbitals \cite{ruiz2019interlayer,cappelluti2013tight,PhysRevResearch.3.043217,MoireExcitonsLinderalv}. Thus we have two periodic modifications to the decoupled Hamiltonian $H_0$ \cite{brem2020tunable,D0NR02160A}. These contributions make up what is in general referred to as moir\'e potential \cite{tong2020interferences,MoireExcitonsLinderalv,D0NR02160A,brem2020tunable,PhysRevResearch.3.043217}. We model this potential by first considering the two underlying effects in real space
\begin{align}\label{eq:RealSpaceHamMoire}
\begin{split}
         &H_M=\sum_{\substack{i\lambda\bm{r}}}V^{\lambda}_{ii}(\bm{r})\Psi^{\lambda\dagger}_i(\bm{r})\Psi^{\lambda}_i(\bm{r})\\
     &+\sum_{\substack{i\neq j\\\lambda\bm{r}}}T^{\lambda}_{ij}(\bm{r})\Psi^{\lambda\dagger}_i(\bm{r})\Psi^{\lambda}_j(\bm{r})+h.c.
    \end{split}
\end{align}
where $i=(l,\xi)$ is a compound index, consisting of the layer index $l$ and valley index $\xi$. Here, $\lambda=(c,v)$ is the band index and $\Psi^{(\dagger)}$ are annihilation(creation) operators. We approximate the wavefunction in the vicinity of high-symmetry points as a plane wave expansion (effective mass approximation) $\Psi^{\lambda\dagger}_i(\bm{r})=\sum_{\bm{k}}e^{i\bm{k}\cdot\bm{r}}\lambda^{\dagger}_{i,\bm{k}}$. The moir\'e Hamiltonian has then the following expression in momentum space
\begin{align}
\begin{split}\label{eq:ElectronHoleHam}
    &H_M=\sum_{\substack{i\lambda\\\bm{k}\bm{g}}}v^{\lambda}_{ii}(\bm{g})\lambda^{\dagger}_{i,\bm{k}+\bm{g}}\lambda_{i,\bm{k}}\\
    &+\sum_{\substack{i\neq j \lambda\\ \bm{g}\bm{k}\bm{q}}}t^{\lambda}_{ij}(\bm{g})\lambda^{\dagger}_{i,\bm{k}+\bm{g}}\lambda_{j,\bm{k}}+h.c.
    \end{split}
\end{align}
Here, the periodic moir\'e potential $V^{\lambda}_{ii}(\bm{r})$ has been expanded as a Fourier series
\begin{equation}\label{eq:Expansion}
    V^{\lambda}_{ii}(\bm{r})=\sum_{\bm{g}}v^{\lambda}_{ii}(\bm{g})e^{i\bm{g}\cdot\bm{r}},
\end{equation}
where $\bm{g}$ are the mini Brilliun zone (mBZ) lattice vectors of the moir\'e superlattice and $v^{\lambda}_{ii}(\bm{g})$ are the Fourier coefficients of the expansion. These can be calculated by solving the integral for the Fourier coefficients
\begin{equation}\label{eq:CoefInt}
        v^{\lambda}_{ii}(\bm{g})=\frac{1}{\mathcal{A}_{\text{M}}}\int_{\mathcal{A}_{\text{M}}}d\bm{r}e^{-i\bm{g}\cdot \bm{r}}V^{\lambda}_{ii}(\bm{r}),
\end{equation}
where $\mathcal{A}_{\text{M}}$ is the unit area of the superlattice. Hence, the matrix element for the moir\'e periodic potentials can be computed via discrete Fourier transforms, if we know the stacking-dependent change in the alignment/tunneling strength. This is done by assuming a rigid lattice model and by smoothly interpolating between high-symmetry points \cite{PhysRevResearch.3.043217,brem2020tunable}
\begin{equation}\label{eq:FitFormula}
    V^{\lambda}_{ii}(\bm{r})=\text{Re}\Big[v^{\lambda}_i+(\mathcal{A}^{\lambda}_i+\mathcal{B}^{\lambda}e^{i2\pi/3})\sum_{n=0}^{2}e^{i\bm{g}_n\cdot\bm{r}}\Big],
\end{equation}
where $\bm{g}_n$ are the $\text{C}_3$-symmetric mBZ lattice vectors in the first shell. The fit parameters $v^{\lambda}_i$, $\mathcal{A}^{\lambda}_i$ and $\mathcal{B}^{\lambda}_i$ are adjusted to match the ab initio results for alignment shifts at a set of high-symmetry stackings computed in Refs. \cite{PhysRevResearch.3.043217,MoireExcitonsLinderalv}. The Fourier coefficients $t^{\lambda}_{ij}(\bm{g})$ for the tunneling follow from the same procedure as the moir\'e induced alignment shift, i.e \autoref{eq:CoefInt} and \autoref{eq:FitFormula}.

Having determined the electronic Hamiltonian for the interlayer coupling effects, we can now derive the changes in exciton energies via the exciton Hamiltonian approach \cite{katsch2018theory}. The following procedure is analogous to the one performed in Ref. \cite{brem2020tunable,D0NR02160A}. 
Assuming low densities, we can expand the electronic operators using pairs $P^{\dagger}_{ik,j\bm{k}^{\prime}}=c^{ \dagger}_{i \bm{k}}v_{j \bm{k}^{\prime}}$, thus yielding the following expression \cite{katsch2018theory,D0NR02160A,brem2020tunable}
\begin{equation}\label{eq:pairoperator}
\begin{split}
c^{\dagger}_{i \bm{k}}c_{j \bm{k}^{\prime}}\approx\sum_{m\bm{p}}P^{\dagger}_{i\bm{k},m\bm{p}}P_{j\bm{k}^{\prime},m\bm{p}}\\
v^{\dagger}_{i \bm{k}}v_{j \bm{k}^{\prime}}\approx\delta^{ij}_{\bm{k}\bm{k}^{\prime}}-\sum_{m\bm{p}}P^{\dagger}_{m\bm{p},j\bm{k}^{\prime}}P_{m\bm{p},i\bm{k}}.
 \end{split}
\end{equation}
Using the pair operator expansion above we can now transform into the exciton basis
\begin{equation}
P^{\dagger}_{i\bm{k},j\bm{k}^{\prime}}=\sum_{\mu}X^{\mu\dagger}_{ij,\bm{k}-\bm{k}^{\prime}+\xi_i-\xi_j}\Psi^{\mu}_{ij}(\alpha_{ij}\bm{k}^{\prime}+\beta_{ij}\bm{k}),
\end{equation}
where $\mu$ is the exciton quantum number, which we restrict to the 1s state in this work. Furthermore, $\Psi^{\mu}_{ij}(\bm{k})$ is the exciton wave function that solves the Wannier equation. Moreover, we have introduced the abbreviation $\alpha_{ij}(\beta_{ij})=m^{c(v)}_{i(j)}/(m^c_i+m^v_j)$. Here, $\xi_i$ is the valley coordinate of the compound $i$. Consequently, we can now write the complete interaction-free exciton Hamiltonian as follows
\begin{equation}\label{eq:ExcitonHam}
\begin{split}
    &H_0=\sum_{L\bm{Q}\bm{\xi}}E^{\bm{\xi}}_{L\bm{Q}}X^{\bm{\xi}\dagger}_{L,\bm{Q}}X^{\bm{\xi}}_{L,\bm{Q}}\\
    &+\sum_{\substack{L\bm{Q}\bm{\xi}\\\bm{g}}}V^{\bm{\xi}}_L(\bm{g})X^{\bm{\xi}\dagger}_{L,\bm{Q}+\bm{g}}X^{\bm{\xi}}_{L,\bm{Q}}\\
    &+\sum_{\substack{LL^{\prime}\\\bm{Q}\bm{\xi}\bm{g}}}T^{\bm{\xi}}_{LL^{\prime}}(\bm{g})X^{\bm{\xi}\dagger}_{L,\bm{Q}+\bm{g}}X^{\bm{\xi}}_{L^{\prime},\bm{Q}}+h.c,
    \end{split}
\end{equation}
where $L=(l_e,l_h)$ is a compound layer index, $\bm{Q}$ is the center-of-mass momentum and $\bm{\xi}=(\xi_e,\xi_h)$ is the valley index. The first term accounts for the dispersion with 
\begin{equation}\label{eq:dispersion}
    E^{\bm{\xi}}_{L\bm{Q}}=\hbar^2\frac{(\bm{Q}-[\xi_e-\xi_h])^2}{2[m_e+m_h]}+\varepsilon^c_{\xi_e0}-\varepsilon^v_{\xi_h0}+E^\text{b}_{\bm{\xi}}.
\end{equation}
Here. $E^\text{b}_{\bm{\xi}}$ are the exciton binding energies and $\varepsilon^{\lambda}_{\bm{\xi}_{\lambda}}$ is the valley splitting. The masses and valley splittings are obtained from Ref. \cite{Korm_nyos_2015}. 

The matrix element of the moir\'e-induced alignment shift is given by
\begin{equation}
\begin{split}
V^{\bm{\xi}}_L(\bm{g})=v^c_{l_{e}}(\bm{g})\mathcal{F}^{\bm{\xi}}_{LL}(\beta_{LL}\bm{g})-v^v_{l_{h}}(\bm{g})\mathcal{F}^{\bm{\xi}*}_{LL}(-\alpha_{LL}\bm{g}),
\end{split}
\end{equation}
where $v^{\lambda}_{l_{\lambda}}(\bm{g})$ are the Fourier coefficients as obtained from \autoref{eq:CoefInt} and $\mathcal{F}^{\bm{\xi}}_{LL^{\prime}}(\bm{q})$ are the form factors given by
\begin{equation}
    \mathcal{F}^{\bm{\xi}}_{LL^{\prime}}(\bm{q})=\sum_{\bm{k}}\Psi^{\bm{\xi}*}_{L}(\bm{k})\Psi^{\bm{\xi}}_{L^{\prime}}(\bm{k}+\bm{q}).
\end{equation}
The tunneling matrix element is similarly given by
\begin{equation}
\begin{split}
T^{\bm{\xi}}_{LL^{\prime}}(\bm{g})=\Big[\delta_{l_h,l_h^{\prime}}(1-\delta_{l_e,l_e^{\prime}})t^{c\bm{\xi}_{e}}_{l_el_e^{\prime}}(\bm{g})\mathcal{F}^{\bm{\xi}}_{LL^{\prime}}(\beta_{LL^{\prime}}\bm{g})\\
-\delta_{l_e,l_e^{\prime}}(1-\delta_{l_h,l_h^{\prime}})t^{v\bm{\xi}_h}_{l_hl_h^{\prime}}(\bm{g})\mathcal{F}^{*\bm{\xi}}_{LL^{\prime}}(-\alpha_{LL^{\prime}}\bm{g})\Big],
\end{split}
\end{equation}
where the delta functions ensure that only an electron or hole can tunnel at the same time and $t^{\lambda\bm{\xi}_{\lambda}}_{l_{\lambda}l_{\lambda^{\prime}}}(\bm{g})$ are the Fourier coefficients of the real space tunneling landscape, in analogy to \autoref{eq:CoefInt}.

\subsection{Application of out-of-plane electrical field}
The Hamiltonian for the additional potential arising from the homogeneous out-of-plane electrical field $E_z$ can be approximated as
\begin{equation}
    H_{field}=-\sum_{\substack{\bm{k}l\\\lambda}}e_0 z^{\lambda}_lE_z\lambda^{\dagger}_{\bm{k}l}\lambda_{\bm{k}l},
\end{equation}
where $e_0$ is the elementary charge and $z^{\lambda}_l$ is the real space position in $z$-direction of electrons (holes) in layer $l$. By using the pair operator expansion defined in \autoref{eq:pairoperator} we can transform this equation into the exciton basis
\begin{equation}
    H_{X-l}=-\sum_{\bm{\xi}\bm{Q}L} d_LE_zX^{\bm{\xi}\dagger}_{\bm{Q},L}X^{\bm{\xi}}_{\bm{Q},L},
\end{equation}
where $d_L=e_0u_L$ is the dipole moment and $u_L$ the dipole length. This can be understood as the potential energy of a dipole within an electrical field and is often referred to as the Stark shift of the exciton. The latter is proportional to the exciton dipole length $u_L=z^c_l-z^v_l$ and the electrical field strength $E_z$. Here the dipole length of an interlayer exciton is taken to be constant due to the small change in the interlayer distance throughout the moir\'e superlattice, where $|d_{\text{iX}}|=0.645$ nm and $|d_{\text{X}_\text{A/B}}|=0.0$ nm \cite{laturia2018dielectric}. The equation above can now easily be incorporated into \autoref{eq:dispersion} for further consideration.

\subsection{Diagonalization of moir\'e exciton Hamiltonian}
In order to diagonalize the Hamiltonian we apply the well-known zone-folding scheme, where we restrict the summation over the center-of-mass momentum $\bm{Q}$ to the mBZ and fold it back with the mBZ lattice vectors $\bm{g}$ \cite{D0NR02160A,brem2020tunable}. Equation (\ref{eq:ExcitonHam}) then becomes
\begin{equation}
\begin{split}
    H_0&=\sum_{\substack{L\bm{Q}\bm{\xi}\\\bm{g}}}E^{\bm{\xi}}_{L\bm{Q}}(\bm{g})X^{\bm{\xi}\dagger}_{L,\bm{Q}+\bm{g}}X^{\bm{\xi}}_{L,\bm{Q}+\bm{g}}\\
    &+\sum_{\substack{L\bm{Q}\bm{\xi}\\\bm{g}\bm{g}^{\prime}}}V^{\bm{\xi}}_L(\bm{g}^{\prime})X^{\bm{\xi}\dagger}_{L,\bm{Q}+\bm{g}+\bm{g}^{\prime}}X^{\bm{\xi}}_{L,\bm{Q}+\bm{g}}\\
    &+\sum_{\substack{LL^{\prime}\bm{Q}\\\bm{\xi}\bm{g}\bm{g}^{\prime}}}T^{\bm{\xi}}_{LL^{\prime}}(\bm{g}^{\prime})X^{\bm{\xi}\dagger}_{L^{\prime},\bm{Q}+\bm{g}+\bm{g}^{\prime}}X^{\bm{\xi}}_{L,\bm{Q}+\bm{g}}+h.c,
       \end{split}
\end{equation}
where $E^{\bm{\xi}}_{L\bm{Q}}(\bm{g})=E^{\bm{\xi}}_{L\bm{Q}+\bm{g}}$ and the summation over $\bm{Q}\in$mBZ. By introducing the zone-folded operator $F_{L\bm{Q}\bm{g}}^{\bm{\xi}}=X^{\bm{\xi}}_{L,\bm{Q}+\bm{g}}$ the Hamiltonian takes the form
\begin{equation}
    \begin{split}
        H_0&=\sum_{\substack{L\bm{Q}\bm{\xi}\\\bm{g}}}E^{\bm{\xi}}_{L\bm{Q}}(\bm{g})F_{L\bm{Q}\bm{g}}^{\bm{\xi}\dagger}F_{L\bm{Q}\bm{g}}^{\bm{\xi}}\\
        &+\sum_{\substack{L\bm{Q}\bm{\xi}\\\bm{g}\bm{g}^{\prime}}}V^{\bm{\xi}}_L(\bm{g},\bm{g}^{\prime})F_{L\bm{Q}\bm{g}^{\prime}}^{\bm{\xi}\dagger}F_{L\bm{Q}\bm{g}}^{\bm{\xi}}\\
        &+\sum_{\substack{LL^{\prime}\bm{Q}\\\bm{\xi}\bm{g}\bm{g}^{\prime}}}T^{\bm{\xi}}_{LL^{\prime}}(\bm{g},\bm{g}^{\prime})F_{L\bm{Q}\bm{g}^{\prime}}^{\bm{\xi}\dagger}F_{L\bm{Q}\bm{g}}^{\bm{\xi}}+h.c,
    \end{split}
\end{equation}
with the abbreviation $T^{\bm{\xi}}_{LL^{\prime}}(\bm{g},\bm{g}^{\prime})=T^{\bm{\xi}}_{LL^{\prime}}(\bm{g}^{\prime}-\bm{g})$. 

A hybrid moir\'e exciton basis is introduced as 
\begin{equation}
    Y^{\dagger}_{\bm{\xi}\eta\bm{Q}}=\sum_{\bm{g}L}\mathcal{C}^{\bm{\xi}\eta*}_{L\bm{g}}(\bm{Q})F^{\bm{\xi}\dagger}_{L\bm{Q}\bm{g}},
\end{equation}
where $\mathcal{C}^{\bm{\xi}\eta*}_{L\bm{g}}(\bm{Q})$ is the mixing coefficient describing the relative contribution between different sub-bands and intra/intralayer excitons. Now, $\eta$ is the new hybrid moir\'e exciton quantum number. Since these mixing coefficients are the eigenvectors of the Hamiltonian they fulfill the following criteria 
\begin{equation}
\begin{split}
    \sum_{L\bm{g}}\mathcal{C}^{\bm{\xi}\eta_1*}_{L\bm{g}}(\bm{Q})\mathcal{C}^{\bm{\xi}\eta_2}_{L\bm{g}}(\bm{Q})=\delta_{\eta_1\eta_2}\\
    \sum_{\eta}\mathcal{C}^{\bm{\xi}\eta*}_{L\bm{g}}(\bm{Q})\mathcal{C}^{\bm{\xi}\eta}_{L^{\prime}\bm{g}^{\prime}}(\bm{Q})=\delta_{LL^{\prime}}\delta_{\bm{g}\bm{g}^{\prime}}.
\end{split}    
\end{equation}
Expanding with the mixing coefficients and summing over $\eta$ yields the following eigenvalue problem
\begin{equation}\label{eq:eigenvalue}
    \begin{split}
        E^{\bm{\xi}}_{L\bm{Q}}(\bm{g})\mathcal{C}^{\bm{\xi}\eta}_{L\bm{g}}(\bm{Q})+\sum_{\bm{g}^{\prime}}V^{\bm{\xi}}_L(\bm{g},\bm{g}^{\prime})\mathcal{C}^{\bm{\xi}\eta}_{L\bm{g}^{\prime}}(\bm{Q})\\
        +\sum_{L^{\prime}\bm{g}^{\prime}}T^{\bm{\xi}}_{LL^{\prime}}(\bm{g},\bm{g}^{\prime})\mathcal{C}^{\bm{\xi}\eta}_{L^{\prime}\bm{g}^{\prime}}(\bm{Q})=\mathcal{E}^{\bm{\xi}}_{\eta\bm{Q}}\mathcal{C}^{\bm{\xi}\eta}_{L\bm{g}}(\bm{Q}).
    \end{split}   
\end{equation}
Solving this numerically gives the final diagonal form of the moir\'e Hamiltonian
\begin{equation}
    H_0=\sum_{\bm{Q}\bm{\xi}\eta}\mathcal{E}^{\bm{\xi}}_{\eta\bm{Q}}Y^{\dagger}_{\bm{\xi}\eta\bm{Q}}Y_{\bm{\xi}\eta\bm{Q}}.
\end{equation}

\subsection{Optical response}
To calculate the photoluminescence (PL) spectra for direct and indirect emission we exploit the phonon-assisted photoluminescence formula derived in Refs. \cite{D0NR02160A,PLBrem}.
\begin{equation}
    \begin{split}
    I_{\sigma}(\omega)\propto\sum_{\bm{\xi}\eta}\frac{|\Tilde{M}^{\bm{\xi}\eta}_{\sigma}|^2}{(\mathcal{E}^{\bm{\xi}}_{\eta\bm{0}}-\hbar\omega)^2+(\gamma^{\bm{\xi}}_{\eta}+\Gamma^{\bm{\xi}}_{\eta})^2}\Big(\gamma^{\bm{\xi}}_{\eta}N^{\bm{\xi}}_{\eta\bm{0}}\\
    +\sum_{\substack{\bm{\xi}^{\prime}\eta^{\prime}\tilde{\bm{q}}\\\pm j}}|\Tilde{G}^{\bm{\xi}^{\prime}\eta^{\prime}j}_{\bm{\xi}\eta,\bm{g}}(\bm{0},\bm{q})|^2N^{\bm{\xi}^{\prime}}_{\eta^{\prime}\bm{\bm{q}}}n^{\pm}_{\Tilde{\bm{q}}}L(\mathcal{E}^{\bm{\xi}^{\prime}}_{\eta^{\prime}\bm{q}}\pm\Omega_{\Tilde{{\bm{q}}}j}-\hbar\omega,\Gamma^{\bm{\xi}^{\prime}}_{\eta^{\prime}})\Big),
    \end{split}   
\end{equation}
where $\omega$ is the photon frequency and $\sigma$ the polarization. The first term in the parenthesis corresponds to the direct light emission. Here, $\gamma^{\bm{\xi}}_{\eta}$ is the radiative width and $N^{\bm{\xi}}_{\eta\bm{Q}}$ the exciton occupation which is approximated by the Boltzmann distribution. This equation can be reduced to the exciton Elliot formula, where $\mathcal{E}^{\bm{\xi}}_{\eta\bm{Q}}$ are moir\'e exciton energies calculated from \autoref{eq:eigenvalue}. Furthermore, $\Gamma^{\bm{\xi}}_{\eta}$ is the non-radiative width, which is approximated as $\Gamma^{\text{KK}}=5.0 \text{ meV}+0.05\cdot10^{-3}\text{ (meV/K) }T$ with the temperature $T$ \cite{brem2019intrinsic,selig2016excitonic,D0NR02160A}. When the K$\Lambda$ exciton is visible, it will be the lowest lying state, therefore $\Gamma^{\text{K}\Lambda}$ is expected to be smaller than $\Gamma^{\text{KK}}$. Here we take $\Gamma^{\text{K}\Lambda}=\Gamma^{\text{KK}}/2$. Furthermore, $\Tilde{M}^{\bm{\xi}\eta}_{\sigma}$ is the optical matrix element, taking into account the oscillator strength of intra- and interlayer excitons. This is given by \cite{D0NR02160A}
\begin{equation}
    \Tilde{M}^{\bm{\xi}\eta}=\sum_{L\bm{g}}M^{L\bm{\xi}}\mathcal{C}^{\bm{\xi}\eta}_{L\bm{g}}(\bm{0})\delta_{\xi_e,\xi_h}\delta_{\bm{g},\bm{0}},
\end{equation}
where $M^{L\bm{\xi}}$ is the oscillator strength given by
\begin{equation}
    M^{L\bm{\xi}}=m_L\frac{1}{2\pi}\sum_{\bm{k}}\Psi_{L}^{\bm{\xi}}(\bm{k}).
\end{equation}
Here $m_{\text{iX}}^2$ is taken to be six orders of magnitude smaller than $m_{\text{X}_\text{A}}^2$ in accordance with Ref. \cite{Merkl2019}. Here, $\eta$ is the initial state and $\bm{\xi}$ the valley index, which is restricted to KK in the resonant case.

The second term corresponds to phonon-assisted recombination, where the exciton scatter with a phonon to a virtual state and then emits a photon. Here $n^{\pm}_{\Tilde{\bm{q}}}=1/2\mp 1/2+n_{\text{B}}(\Omega_{\tilde{\bm{q}}j})$, where $n_{\text{B}}(\Omega_{\tilde{\bm{q}}j})$ is the Bose-Einstein distribution and the sum over +(-) corresponds to phonon absorption (emission). Furthermore, $L(\mathcal{E}^{\bm{\xi}^{\prime}}_{\eta^{\prime}\bm{q}}\pm\Omega_{\Tilde{{\bm{q}}}j}-\hbar\omega,\Gamma^{\bm{\xi}^{\prime}}_{\eta^{\prime}})$ is the Cauchy-Lorentz distribution for the final moir\'e exciton energy $\mathcal{E}^{\bm{\xi}^{\prime}}_{\eta^{\prime}\bm{q}}$, which is shifted by a phonon energy $\Omega_{\Tilde{{\bm{q}}}j}$. The energy of optical phonons is taken constant (Einstein model) and the long-range acoustic phonons are approximated as linear in momentum (Debye model). Here $j=(\zeta_j,l_{ph})$ is a compound index, where $\zeta_j$ is the phonon mode index and $l_{ph}$ is the phonon layer index. Furthermore, $\Tilde{\bm{q}}=\bm{q}+\bm{g}$ is the zone-folded phonon momentum. The exciton-phonon matrix element $\Tilde{G}^{\bm{\xi}^{\prime}\eta^{\prime}j}_{\bm{\xi}\eta,\bm{g}}(\bm{Q},\bm{q})$ is given by
\begin{equation}
    \begin{split}
    \Tilde{G}^{\bm{\xi}^{\prime}\eta^{\prime}j}_{\bm{\xi}\eta,\bm{g}}(\bm{Q},\bm{q})=\sum_{\substack{LL^{\prime}\\\bm{g}^{\prime}\bm{g}^{\dprime}}}\Big(G^{c,\bm{\xi}^{\prime}L^{\prime}}_{j,\bm{\xi}L}(\Tilde{\bm{q}}-\Delta Q_{\bm{\xi}\bm{\xi}^{\prime}})\delta_{\xi_h,\xi^{\prime}_h}\delta_{l_e,l_{ph}}\\
    -G^{v,\bm{\xi}^{\prime}L^{\prime}}_{j,\bm{\xi}L}(\Tilde{\bm{q}}-\Delta Q_{\bm{\xi}\bm{\xi}^{\prime}})\delta_{\xi_e,\xi^{\prime}_e}\delta_{l_h,l_{ph}}\Big)\\
    \times\mathcal{C}^{\bm{\xi}\eta*}_{L\bm{g}^{\prime}}(\bm{Q})\mathcal{C}^{\bm{\xi^{\prime}}\eta^{\prime}}_{L^{\prime}\bm{g}^{\dprime}}(\bm{Q}+\bm{q})\delta_{L,L^{\prime}}\delta_{\bm{g}^{\dprime}-\bm{g}^{\prime},\bm{g}},
    \end{split}   
\end{equation}
where $\delta_{\xi_{\lambda},\xi^{\prime}_{\lambda}}$ ensure that we only have single phonon processes and $\delta_{l_{\lambda},l_{ph}}$ enforces that the phonon needs to be in the same layer as the carrier it scatters with. The selection rule $\delta_{L,L^{\prime}}$ restricts the phonons to one layer, i.e. we assume that there is no phonon-tunneling. Since the in-plane atomic forces are far stronger than the interlayer van der Waals interaction, this approximation holds. The matrix element is evaluated in local valley coordinates where $\Delta Q_{\bm{\xi}\bm{\xi}^{\prime}}=\xi^{\prime}_e-\xi^{\prime}_h-\xi_e+\xi_h$. Furthermore, $G^{c,\bm{\xi}^{\prime}L^{\prime}}_{j,\bm{\xi}L}(\bm{q})=g^{c}_{j,\bm{\xi}\bm{\xi}^{\prime}\bm{q}}\mathcal{F}_{L\bm{\xi},L^{\prime}\bm{\xi}^{\prime}}(\beta\bm_{LL^{\prime}}{\bm{q}})$ and $G^{v,\bm{\xi}^{\prime}L^{\prime}}_{j,\bm{\xi}L}(\bm{q})=g^{v}_{j,\bm{\xi}\bm{\xi}^{\prime}\bm{q}}\mathcal{F}_{L\bm{\xi},L^{\prime}\bm{\xi}^{\prime}}(-\alpha\bm_{LL^{\prime}}{\bm{q}})$, where the coupling $g^{\lambda}_{j,\bm{\xi}\bm{\xi}^{\prime}\bm{q}}$ is given by
$
    g^{\lambda}_{j,\bm{\xi}\bm{\xi}^{\prime}\bm{q}}=\mathcal{D}^{\lambda}_{j,\bm{\xi}\bm{\xi}^{\prime}}(\bm{q})\sqrt{\frac{\hbar^2}{2\rho\mathcal{A}\hbar\Omega_{j\bm{q}}}},
$
where $\rho$ is the unit mass density and $\mathcal{A}$ is the associated area of the density. Here $\mathcal{D}^{\lambda}_{j,\bm{\xi}\bm{\xi}^{\prime}}(\bm{q})$ is the electron-phonon coupling deformation potential as obtained from \textit{ab initio} calculations in Ref. \cite{jin2014intrinsic}. 


%